\documentclass[10pt]{article}

\usepackage{amsmath}
\usepackage{amssymb}

\usepackage{graphicx}

\usepackage{cite}
\usepackage{color} 

\usepackage{setspace} 

\usepackage[labelfont=bf,labelsep=period,justification=raggedright]{caption}

\makeatletter
\renewcommand{\@biblabel}[1]{\quad#1.}
\makeatother

\date{}

\newcommand{\gene}[1]{\emph{#1}}
\newcommand{\base}[1]{\texttt{#1}}

\newcommand{\roughselection}{$0.8$\%~}
\newcommand{\recrate}{1.4\times 10^{-5}}
\newcommand{\recrateerr}{1.4\pm 0.6\times 10^{-5}}

\begin{document}

\begin{flushleft}
{\Large
\textbf{Recombination rate and selection strength in HIV intra-patient evolution}
}
\\
Richard A.~Neher$^{1,\ast}$, 
Thomas Leitner$^{2}$
\\
\bf{1} Kavli Institute for Theoretical Physics, University of
California, Santa Barbara, CA 91306
\\
\bf{2} Theoretical Biology and Biophysics, Los Alamos National Laboratory, Los
Alamos, NM 87545
\\
$\ast$ E-mail: neher@kitp.ucsb.edu
\end{flushleft}

\section*{Abstract} 
The evolutionary dynamics of HIV during the chronic phase of
infection is driven by the host immune response and by selective pressures
exerted through drug treatment. To understand and model the evolution of HIV
quantitatively, the parameters governing genetic diversification and the strength
of selection need to be known. While mutation rates can be measured in single
replication cycles, the relevant effective recombination rate depends on the
probability of coinfection of a cell with more than one virus and can only be
inferred from population data. However, most population genetic estimators for
recombination rates assume absence of selection and are hence of limited
applicability to HIV, since positive and purifying selection are important in
HIV evolution. Yet, little is known about the distribution of selection differentials 
between individual viruses and the impact of single polymorphisms on viral
fitness.

Here, we estimate the rate of recombination and the distribution of selection
coefficients from time-resolved sequence data tracking the evolution of HIV
within single patients. By examining temporal changes in the genetic composition
of the population, we estimate the effective recombination to be 
$\rho=\recrateerr$ recombinations per site and generation.
Furthermore, we provide evidence that for at least 15\% of the non-synonymous
polymorphisms observed, selection coefficients exceed \roughselection
per generation.

These results provide a basis for a more detailed understanding of the
evolution of HIV. A particularly interesting case is evolution in response to
drug treatment, where recombination can facilitate the rapid acquisition of multiple
resistance mutations. With the methods developed here,  more precise and more
detailed studies will be possible, as soon as data with higher time resolution
and greater sample sizes is available.

\section*{Author Summary} Evolution, in viruses and other organisms, is the
result of random genetic diversification by mutation and recombination, selection as
well as the stochastic nature of replication and death. For most organisms, these
evolutionary forces have to be estimated from static snapshots of the population.
This inference requires models of the population dynamics that typically assume
that some of the evolutionary forces can be neglected. In rapidly and vigorously
evolving organisms like HIV, such assumptions are questionable. As opposed to
most other cases, however, time resolved data is available for HIV. We present a
direct estimation of the recombination rate and selection coefficients
in the intrapatient dynamics of HIV using time resolved data. We find that the
effective recombination rate in HIV is similar to the substitution rate at
$\rho=\recrateerr$ per generation and site, rather than much larger as
previously reported. Furthermore, we study the distribution of selection coefficients and
find that a significant fraction (15\%) of the observed non-synonymous
polymorphisms in the \gene{env} are selected for with coefficients ranging from
\roughselection to 2\% per generation. Currently, the estimation is limited by the
scarceness of the available data, but much more detailed and accurate studies will be possible in
the near future.

\section*{Introduction} The human immunodeficiency virus (HIV-1) ranks among the
most rapidly evolving entities known \cite{Duffy:2008p26429}, enabling the virus
to continually escape the immune system. After infection with HIV, patients
typically enter an asymptomatic period lasting several years during which the
virus is present at low to medium levels, typically at a viral load of $<50$ to
$10^{4}$ copies per ml plasma. Nevertheless, the number of virions produced and
removed is estimated to be around $10^{10}$ per day with a generation time slightly less
than two days \cite{Perelson:1996p23158}. Due to this rapid turnover and the high
mutation rate of $2.5\times 10^{-5}$ per site and generation, the sequence
diversity of HIV within a single patient can rise to $\approx 5$\% (\gene{env}
gene) within a few years and the divergence from the founder strain increases by
$\approx 1$\% per year \cite{Shankarappa:1999p20227}, although this rate is not
constant \cite{Lee:2008p25480}. The genotypic diversity is subject to positive
selection for novel variants that are not recognized  by the host immune
system or that reduce the sensitivity to anti-retroviral drugs \cite{Nielsen:1998p25205,Chen:2004p22606,Bazykin:2006p17919}, as well as to
purifying selection by functional constraints \cite{Edwards:2006p20328}. In
addition to high substitution rates and strong selection, genomes of different
HIV particles within the same host frequently exchange genetic information. This
form of viral recombination works as follows: Whenever a cell is coinfected by
two or more viruses, the daughter virions can contain two RNA strands from
different viruses\cite{Jung:2002p23469,Chen:2009p27652}. In the next round of
infection, recombinant genomes are generated by template switching of the retrotranscriptase while producing cDNA.
It has been shown that recombination in HIV contributes significantly to the
genetic diversity within a patient
\cite{Liu:2002p20788,Charpentier:2006p20791,Shriner:2004p20830}. In cases
of super-infection with several HIV-1 subtypes, recombination can give rise to
novel forms that become part of the global epidemic \cite{LANLSequence}.

The observation of recombinant viruses after a change in
anti-retroviral drug therapy \cite{Nora:2007p20790} suggests that
recombination might play an important role in the evolution of drug
resistance, as predicted by theoretical models
\cite{Rouzine:2005p17398}. The amount by which recombination speeds up the
evolution of drug resistance depends on the parameters governing the population dynamics
\cite{Althaus:2005p24460}, many of which are not known to sufficient accuracy.
\emph{In vitro} estimates of the recombination rate have shown that the
retrotranscriptase switches templates about $2-3$ times while transcribing the
entire genome, resulting in a recombination rate of $3\times 10^{-4}$ per site
and generation \cite{Jetzt:2000p23004,Zhuang:2002p23003}. However, the bare
template switching rate is only of secondary importance, since recombination can
generate diversity only if the virion contains two RNA strands that originate
from different viruses, which requires coinfection of host
cells\cite{Levy:2004p23309}. The effective \emph{in vivo} recombination rate is
therefore a compound quantity, to which the template switching rate and the
probability of coinfection of a single host cell contribute. This effective
recombination rate has been estimated with coalescent based methods developed in
population genetics \cite{McVean:2002p20849,Shriner:2004p20830}. These methods
use a single sample of sequences obtained from the diverse population and
estimate the recombination rate from topological incongruencies in the
phylogenetic tree of the sequence sample. Together with an estimate of the
mutation rate, this allows to estimate the recombination rate in real time units.
Shriner et al.~\cite{Shriner:2004p20830} report an estimate of $1.38\times
10^{-4}$ per site and generation, implying almost ubiquitous coinfection of host
cells. 

Here, we present a different method to estimate recombination rates from longitudinal
sequence data, which has been obtained from 11 patients at approximately 6 month intervals
\cite{Kaslow:1987p23166,Shankarappa:1999p20227}. By comparing sequence samples from successive
time points, we can estimate recombination rates from the distance and time
dependence of the probability of cross-over between pairs of polymorphic sites.
We find that the effective rate of recombination is $\rho \approx \recrate$
per site and generation. 
Furthermore, we estimate the strength of selection on
nonsynonymous polymorphisms by measuring the rate at which allele frequencies
change. We find that a fraction of about 15\% of the observed nonsynonymous
polymorphisms are selected stronger than \roughselection {} per generation. 	

\begin{figure}[htp]
\begin{center}
  \includegraphics[width=0.5\columnwidth]{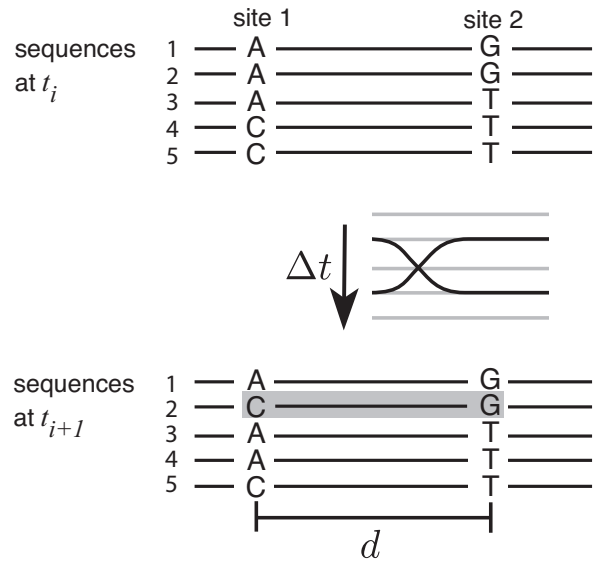}
  \caption{{\bf Recombination produces new haplotypes.} 
  At time point $t_i$, a pair of polymorphic sites is observed in three different combinations 
  \base{A\ldots G}, \base{A\ldots T}, and \base{C\ldots T}. Recombination can generate 
  the missing haplotype \base{C\ldots G} (grey box) from one time point to the
  next with a rate proportional to the product of the distance $d$ between the 
  two sites and the time interval $\Delta t$ between the two  observations. 
  Since no other process depends on 
  this combination of parameters, we can estimate the
  recombination rate using this dependence.}
  \label{fig:hapgeneration}
\end{center}
\end{figure}

\section*{Results}
Time resolved sequence data of the C2-V5 region of the \gene{env} gene was
available from eleven patients over a time-span of 6-13 years with sequence
samples taken typically every 6-10 month \cite{Shankarappa:1999p20227,Shriner:2004p22998}. 
At each time point, about 5-20 viruses had been sequenced successfully. 
These sequences were aligned as described in Methods. We will use the temporal
variation in genetic diversity to estimate the recombination rate and
selection coefficients that govern the dynamics of the viral population.

\subsection*{Recombination rate} Recombination produces new combinations of
alleles from existing genetic variation and randomizes the distribution of
genotypes. To illustrate this process and the challenges of estimating
recombination rates, consider the pair of polymorphic sites in Figure
\ref{fig:hapgeneration}. 
Generically such a pair will have arisen
by the following sequence of events: (i) Site 1 becomes polymorphic by mutation,
e.g. \base{A}/\base{C}. (ii) A mutation occurs at site 2 on a genome that carries
one of the variants of site 1, e.g. giving rise to the haplotypes \base{A\ldots
T}, \base{A\ldots G} and \base{C\ldots T}. (iii) The missing haplotype, e.g.
\base{C\ldots G}, can be generated by further mutation (\base{A}$\to$\base{C} at
site 1 or \base{T}$\to$\base{G} at site 2) or by crossing over two of the
existing haplotypes, as illustrated in Figure~\ref{fig:hapgeneration}. In
population genetics, the occurrence of the fourth haplotype is often taken as
sufficient condition for recombination (the four gamete
test\cite{Hudson:1985p23392}). While this is true for bacteria and eukaryotes
because of their low mutation rates, the HIV population within a patient
is large and mutates rapidly \cite{Leitner:1997p23019}. Hence, the biggest challenge
in estimating recombination rates is to separate recombination from recurrent
mutations or homoplasy. A second confounding effect stems from the small number
of sequences available per time point, such that the sequences containing the
fourth haplotype could have been missed due to undersampling. To disentangle
these two effects from recombination, we make use of the fact that only the
recombination rate depends strongly on the distance between the two sites, while
recurrent mutations and sampling noise should not. 

\begin{figure}[htp]
\begin{center}
  \includegraphics[width=0.5\columnwidth]{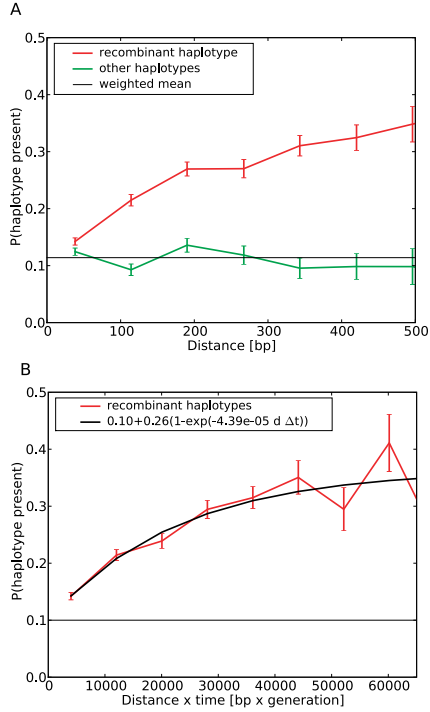}  
  \caption{{\bf Estimating recombination rates from time resolved data.} 
  Panel A shows the probability of finding a haplotype that is not detected at time 
  $t_i$ in the next time slice $t_{i+1}$ as a function of the separation $d$ of
  the sites. The data labeled `recombinant haplotypes' refers to those combinations, that can 
  be generated by recombination from the alleles detected at time $t_i$,
  and displays a pronounced distance dependence. The data labeled `other haplotypes' refers 
  to pairs containing at least one allele not detected at time $t_i$,
  implying an additional mutation or undersampling. The data is averaged over all 
  time points, patients and those pairs of polymorphic sites, where
  both alleles at both sites are seen at least 3 times. 
  Panel B shows the probability of finding the missing 
  haplotype as a function of the product of distance $d$ and time interval
  $\Delta t= t_{i+1}-t_i$. The fit to the data
  is shown in black with fit parameters indicated in the legend. }
  \label{fig:recombination_rate}
\end{center}
\end{figure}

For each pair of biallic sites at time $t_i$ that was found in three of the
four possible haplotypes, we asked whether the missing haplotype is observed
the at time $t_{i+1}$ (comp.~figure \ref{fig:hapgeneration}) and calculated the frequency of
this event as a function of the separation of the two sites, as shown in figure
\ref{fig:recombination_rate}A. This frequency increases with the separation of the two sites
from about 0.1 to about 0.35 at 500 bp separation, in line 
with the expectation that recombination is more rapid between distant sites. To
corroborate that this distance dependence is indeed due to recombination, we
performed the following similar analysis: The curve labelled ``other haplotypes'' in
figure \ref{fig:recombination_rate}A shows the frequency of observing a
haplotype at time $t_{i+1}$, which contains alleles not present at time $t_i$
again averaged over all available data. Any such haplotype could have arisen by
mutation in the time interval between $t_i$ and $t_{i+1}$, or could have been present
at time $t_i$ but not sampled. It cannot, however, be assembled by recombination
from the alleles found at time $t_i$. The important observation is that the frequency of
observing such a haplotype does not increase with distance. This is consistent
with our expectation that an additional mutation or undersampling should be
independent of an additional polymorphism nearby. The clear separation
between the two classes of haplotypes suggests that the contribution from
homoplasy and sampling can be accounted for by a distance independent constant.

\begin{figure}[htp]
\begin{center}
  \includegraphics[width=0.5\columnwidth]{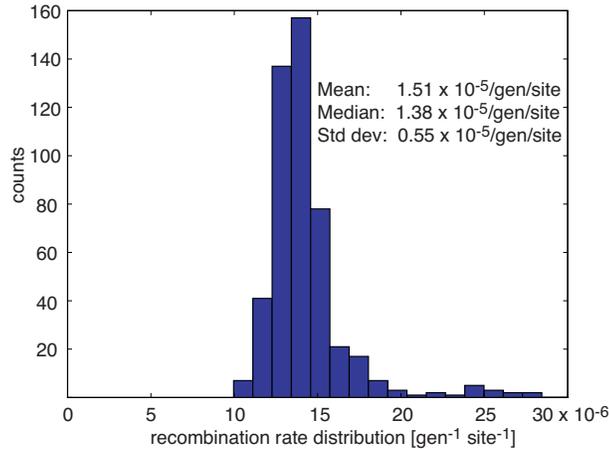}
  \caption[]{{\bf Bootstrap distribution of the recombination rate estimate.}
  The variability of the recombination rate estimator was assessed by repeating the
  fit 500 times to data of eleven patients sampled with replacement from the
  original eleven patients. The mean, median and 
  standard deviation of the distribution are given in the figure. }
  \label{fig:recratedis}
\end{center}
\end{figure}

The probability of recombination between two sites increases with the
product of the time elapsed and the distance between the sites, rather than
with distance alone as plotted in figure \ref{fig:recombination_rate}A. Panel B of figure
\ref{fig:recombination_rate} shows the probability of appearance of a putative 
recombinant haplotype as a function of the product of distance and number 
of generations (generation time 2 days). To estimate the recombination rate,
we have to know how the observed saturation behavior is related to 
recombination. Let $a/A$ and $b/B$ be the alleles at site 1 and 2 and $p_{AB}$
the frequency of the missing haplotype $AB$. The probability not to detect
haplotype $AB$ in a sample of size $M$ is 
\begin{equation}
P(\mathrm{no\;AB})=(1-p_{AB})^M\approx e^{-Mp_{AB}}
\end{equation}
Assuming the allele frequencies remain constant, $p_{AB}$
will relax from its initial value $p_{AB}^0$ to the product of the allele
frequencies $p_Ap_B$ as $p_{AB}(\Delta t)=p_Ap_B + (p_{AB}^0-p_Ap_B)e^{-\rho
d\Delta t }$ through recombination\cite{Ewens2004}. The frequency of detecting
a haplotype at time $t_{i+1}$, given it was not detected in the previous sample at time $t=t_{i}$,
is therefore ${\cal P}(d (t_{i+1}-t_{i}))=1- \langle
e^{-Mp_{AB}(t_{i+1})}\rangle$. This quantity, averaged over all pairs of
polymorphic sites at distance $d$ such that $d (t_{i+1}-t_{i})$ falls into a
specified bin, is shown in figure \ref{fig:recombination_rate}B (the average
extends over all patients and time points). To understand how ${\cal P}(d
(t_{i+1}-t_{i}))$ depends on the recombination rate, it is useful to consider the two limiting cases of small and large $d\Delta t$
\begin{equation}
\label{eq:prob_finding}
{\cal P}(d \Delta t)=\begin{cases}
           \langle Mp_{AB}^0\rangle +\langle
           M(p_Ap_B-p_{AB}^0)\rangle \rho d\Delta t & \rho d\Delta t \ll 1
           \\ 1-\langle e^{-Mp_Ap_B}\rangle & \rho d\Delta t  \gg 1
           \end{cases}
\end{equation}
where we assumed that $\langle M p_{AB}^0\rangle$ is small
compared to one (inspection of figure \ref{fig:recombination_rate}B shows this
is true, we discuss this in more detail in Methods).
Hence, the recombination rate is proportional to the slope $\alpha$ of ${\cal P}(d\Delta t)$ at $d\Delta t=0$.
The intercept ${\cal P}(0)=\langle Mp_{AB}^0\rangle$ is simply the
probability that we detect $AB$ at time $t_{i+1}$ in absence 
of recombination, given we missed it at time $t_{i}$. We determine the slope
and the intercept by fitting the function $f(d\Delta t)=c_0+c_1 (1-e^{-c_2 d\Delta t})$ 
to the data. The recombination rate is then given by
\begin{equation}
\hat{\rho}=\frac{c_1 c_2}{\langle M p_A p_B\rangle-c_0},
\end{equation}
where $\langle M p_A p_B\rangle \approx 0.93$ is measured directly.  
The best fit yields $\hat{\rho}\approx \recrateerr$ recombinations per
site and generation ($\pm$ one standard deviation). The uncertainty of $\hat{\rho}$
was estimated by resampling the patients with replacement 500 times. This bootstrap 
distribution of the recombination rate estimate is shown in figure
\ref{fig:recratedis}. We have assumed that the allele frequencies $p_A$ and
$p_B$ remain constant in the interval $[t_{i+1}, t_i]$. We will see
below, however, that some allele frequencies change rapidly. We expect that
repeated sweeps will cause our method to overestimate the recombination rate:
When the frequencies of the minor alleles increase, the missing haplotype is
produced more rapidly then expected. 

\subsection*{Selection} Positive selection on the variable regions in \gene{env}
and purifying selection on the conserved regions have been repeatedly reported in
the literature
\cite{Seibert:1995p23263,Nielsen:1998p25205,Templeton:2004p23262,Yamaguchi:1997p23295,Shriner:2004p22998}.
Most of the these studies compared the rates of synonymous and non-synonymous
substitutions ($dn/ds$ ratio). An observation of $dn/ds>1$ indicates selection
for novel variants, while $dn/ds<1$ indicates that the amino acid sequence
changes more slowly than the nucleotide sequence, indicating functional
constraint at the protein level. The overall rate of substitutions, however,
yields only very limited information about the strength of selection.

Selection for a specific variant implies that this variant confers an elevated
rate of reproduction compared to the population mean. Ignoring random drift, the
strength of selection $s$ is related to the rate at which the frequency $p$ of 
the variant changes \cite{Ewens2004}:
\begin{equation}
\label{eq:afchange}
\frac{d p}{d t} = p(1-p) s
\end{equation}
This equation has a straightforward solution $p(t)=e^{st}/(1+e^{st})$ and
from two measurements of $p(t)$ at time $t_i$ and $t_{i+1}$, $s$ can be
determined to be $s=\frac{1}{t_{i+1}-t_i}\ln
\frac{p(t_i)q(t_{i+1})}{q(t_i)p(t_{i+1})}$ ($q=1-p$)\cite{Liu:2002p20788} 
(for a review on selection and drift, see \cite{Rouzine:2001p26428}). However,
when using this formula, emphasis is put on rare alleles, whose frequencies can't 
be measured accurately in small samples. It proves more
robust to estimate the rate of allele frequency change directly as $\frac{\Delta
p}{\Delta t}$, where $\Delta p$ is the difference in allele frequency between
consecutive samples and $\Delta t$ is the time interval. The discrete
derivative $\frac{\Delta p}{\Delta t}$ can serve as a proxy for selection which
is less sensitive to rare alleles. The observed $\Delta p$ will be a sum of
contributions from selection, noise from random sampling, and genetic drift. The
latter can be estimated by measuring $\Delta p$ for synonymous polymorphisms which are putatively
neutral and their observed frequency changes are assumed to be dominated by
sampling noise. However, they can be affected by selection on nearby
non-synonymous polymorphisms (hitch-hiking) or be themself under selection,
e.g.~for translation efficiency or RNA secondary structure. 

\begin{figure}[htp]
\begin{center}
  \includegraphics[width=0.5\columnwidth]{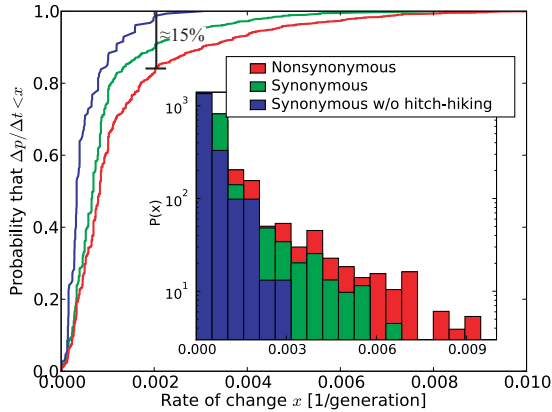}  
  \caption[]{{\bf Selection on non-synonymous mutations.} The figure shows cumulative distributions 
  of the observed rate of change $\Delta p/\Delta t$ of the allele frequencies $p$ between two
  consecutive time points $t_i$  and $t_{i+1}$ for non-synonymous
  polymorphisms, synonymous polymorphisms and synonymous polymorphisms at
  least 100bp away from the nearest fast changing non-synonymous 
  polymorphism ($>0.002$/generation), see methods. The insets show the
  corresponding histograms on a logarithmic scale. 
  Only pairs of time points with sample sizes greater
  than 10 sequences are included. }
  \label{fig:selection_strength}
\end{center}
\end{figure}

Despite the limited resolution of $\Delta p$ and  $\Delta t$ due to small sample
sizes and large time intervals between samples (6-10 month), we can make a
meaningful statement about the strength of selection when averaging over all sites, patients
and time points. Figure \ref{fig:selection_strength} shows the cumulative
distributions of the rates of change of allele frequencies observed during the
interval between two consecutive time points for non-synonymous and synonymous
polymorphisms. The histograms are shown as insets in the figure. There are
consistently more fast changing non-synonymous polymorphisms than there are
synonymous ones, suggesting that a fraction of the non-synonymous polymorphisms
is indeed responding to selection. To check whether the fast changing synonymous
polymorphisms can be attributed to hitchhiking, we excluded synonymous
polymorphisms that are closer than 100bp to a non-synonymous polymorphism that
changes faster than $0.002$ per generation. The resulting distribution is much
narrower with no allele frequency changes beyond $0.003$ per generation,
indicating that the fast changing synonymous polymorphisms are indeed 
``hitch-hiking''. The cumulative histograms can be compared by the
Kolmogorov-Smirnov test, which uses the maximal vertical distance between curves
as a test statistics. The test reveals that the non-synonymous
distribution is significantly different from both the unconditional
synonymous distribution (p-value $<10^{-13}$) and the synonymous distribution
without hitch-hiking (p-value $<10^{-25}$). Note that not all observations are
independent since nearby sites are linked and move coherently. Hence,
realistic p-values will be larger.

The fastest allele frequency changes detected are about $0.01$ per generation,
which is our resolution limit (6 month $\sim$ 100 generations). This low time resolution
results in a tendency to underestimate the rates of change, while the finite
sample size will generate spurious frequency changes due to sampling noise. 
However, the narrow distribution of
allele frequency changes of synonymous polymorphisms excluding hitchkiking
suggests that the contribution from sampling noise is small (on the order of
$0.001$ per generation). Hence, the tail of the histogram of the frequency changes of nonsynonymous
polymorphisms contains a rough measure of the distribution of selection
coefficients\cite{AdamEyreWalker:2007p24636}: about 15\% of the observed
non-synonymous polymorphisms change faster than $0.002$ per generation, compared 
to almost none of the synonymous polymorphisms. Assuming that this difference
is due to selection, we conclude that about 15\% of the observed non-synonymous polymorphisms are
positively selected with $s> 8\times 10^{-3}$ per generation. In the last step,
we rearranged Eq.~(\ref{eq:afchange}) to obtain 
$s \approx \frac{1}{p(1-p)} \frac{\Delta p}{\Delta t} \geq 4 \times \frac{\Delta p}{\Delta t}$ 
with $p(1-p)\leq 0.25$. Most of the strongly deleterious polymorphism are of
course never observed in the samples. 

\section*{Discussion} The dynamics of HIV within a single patient is
characterized by large diversity due to high mutation rates, intense selection,
frequent recombination and stochasticity resulting from bursts of viruses
originating from a single cell. The simultaneous importance of these four
evolutionary forces makes HIV evolution difficult to analyze with traditional
population genetic methods, which typically assume that one of the evolutionary
forces is predominant. Coalescent based methods, for example, assume that
evolution is neutral, i.e. stochasticity dominates over selection. While it is
possible to include recombination or selection into a coalescent description
\cite{Griffiths:1996p25579,Neuhauser:1997p19663}, the analysis becomes difficult
and computationally demanding. Estimators based on site frequency spectra, like
Tajima's $D$, work best when recombination is strong compared to selection.
Phylogenetic analysis, on the other hand, assumes absence of recombination. These
methods have been designed to infer parameters of the population dynamics from
snapshots of the population, which is a formidable challenge. The lack of time
resolved data requires the assumption of a model of the population dynamics,
which can be extrapolated back in time to the most recent common ancestor. 
In neutral models, the time to the  most recent common ancestor is on
the order of $N$ generations, $N$ being the population size. During this
long time interval, there is ample opportunity for selection or demography to
invalidate the assumption of the model.

If time resolved data is available, the task of estimating parameters is greatly
simplified since one can trace the dynamics of alleles and genotypes directly.
Such longitudinally sampled data has for example been used to gauge the molecular
clock of bacterial evolution \cite{Wilson:2009p27988}. We have used time resolved
data from 11 patients to estimate parameters of the population dynamics directly.
In accordance with existing studies, we find that recombination in HIV is
frequent and contributes significantly to sequence diversity
\cite{Shriner:2004p20830,Liu:2002p20788,McVean:2002p20849,Charpentier:2006p20791}.
The template switching rate of HIV is known to be about $3\times 10^{-4}$ per
site\cite{Jetzt:2000p23004,Zhuang:2002p23003}. For template switching to result
in a novel genetic variant, the two RNA strands in the infecting virus have to be
different, which implies co-infection of the cell the virus originated from. Any
estimate of recombination rates from sequence diversity will therefore measure an
effective recombination rate $\rho$ being approximately the product of the
coinfection probability, the probability of forming a heterozygote
and the template switching rate. Our estimate of this effective recombination rate 
is $\rho\approx \recrate$ recombination per site and
generation, which is about a factor of 20 lower than the template switching rate
but still implies a probability of coinfection of about 10\%. Our estimate is
lower than the estimate by Shriner et al.~\cite{Shriner:2004p20830}, who estimated
$\rho\approx 1.38\times 10^{-4}$ per site and generation. However, two other
estimators also reported in that paper yielded lower recombination rates
comparable to our estimate.

In our analysis we have assumed that the rate of recombination is constant
across the \gene{env} gene. However, the breakpoint distribution in circulating
recombinant forms show strong variation along the genome, with particulary
little recombination in \gene{env}\cite{Archer:2008p17710}. The variation of
the breakpoint distribution can largely be explained by low sequence homology
between different subtypes and dysfunctional recombinants \cite{SimonLoriere:2009p25916}. 
In the present study, however, all patients were infected with a single subtype
and gradually built up diversity which remained much lower than the distance
between subtypes. The effects causing recombination rate variation
should therefore be of minor importance.

By comparing the distribution of the rate of allele frequency changes of
synonymous (putatively neutral) and nonsynonymous (possibly selected)
polymorphisms, we estimated the distribution of selection coefficients on single
sites. We find that 15\% of the nonsynonymous polymorphism are selected with
coefficients greater than \roughselection {}per generation. In using the
distribution of allele frequency changes to infer selection coefficients, we have
assumed that each locus is selected for its own effect on fitness, rather than
changing its frequency due to selection on a linked neighboring locus
\cite{Gillespie:2001p9636} or some epistatic combination of loci
\cite{Kimura:1965p3008,Neher:2009p22302}. A sweeping polymorphism ``drags'' along
neutral variation in a region $s/\rho\log Ns$
\cite{Barton:1998p28270}, which using our estimates of $\rho$, $s\approx
0.01$ and an effective population size of $10^4$ evaluates to $200$bp. This is
consistent with our finding that most of the rapid allele frequency changes 
of synonymous polymorphisms occur in the vicinity ($<100bp$) of rapidly
changing non-synonymous polymorphisms (figure \ref{fig:selection_strength}).

The sequence diversity in our samples is on the order of 3\%
\cite{Shankarappa:1999p20227}, such that two polymorphisms are expected to be on
average 30bp apart. Roughly half of the observed polymorphisms are
non-synonymous, of which 15\% are under strong positive selection. Hence the
distance between simultaneously sweeping loci is on the order of 400bp, 
which is of the same order as $s/\rho$. If the rate of sweeps was much
higher, sweeps would cease to be independent and interfere. While these 
estimates have large uncertainty, it is conceivable that the rate of sweeps is
limited by recombination.

Selection coefficients in HIV have been estimated before by Liu et
al.~\cite{Liu:2002p20788} in a patient infected with two HIV-1 subtype B
viruses. In this patient, a small number of recombinant forms competed against
the ancestral strains and selection differentials were estimated to lie between
$0.3$\% and $9$\%. These selection coefficients are higher than our estimates,
which is plausible since they are associated with entire recombinant genomes that
differ at many sites rather than the single site estimate presented here.
The strength of selection for novel epitopes in several HIV genes (rate
of escape from cytotoxic T-lymphocyte mediated killing) during the asymptomatic
phase of HIV infection has been shown to be of similar magnitude as our
estimates \cite{Asquith:2006p28003}.

The present study is limited by low time resolution and small sequence samples
and more accurate and detailed answers could be obtained from larger samples.
Larger sample sizes will require generalizations of the method used to estimate
recombination rates which depends on pairs of sites where one of the four
possible haplotypes is absent. In large samples, high frequency variants will
most likely be present in all four possible haplotypes. In this case, one can
measure linkage disequilibrium $D$ directly and observe how it decays from one
timepoint  to the next, e.g.~by measuring its autocorrelation function $\langle
D(t_i)D(t_{i+1}) \rangle = \langle D^2 \rangle e^{-\rho d\Delta t}$. The method
presented here is an approximation of this more general method suiteable for
noisy data.

\section*{Methods}
\subsection*{Longitudinal sequence data and alignment} Sequences of the C2-V5
region of \gene{env} from 11 patients, which were part of the MACS
study \cite{Kaslow:1987p23166,Shankarappa:1999p20227}, were obtained from the
Los Alamos National Lab HIV data base 
(Special interest alignments, accession numbers: AF137629-AF138703, AF204402-AF204670, AY449806 -
AY450055 \& AY450056 - AY450257). The sequences were translated into amino acid 
sequences and aligned for each patient separately using MUSCLE v3.6 
with default settings \cite{Edgar:2004p23254}. Aligned nucleotide sequences 
were then constructed by inserting a 3bp gap for each  gap in the amino acid
alignment. Scripting and plotting was done in Python using the NumPy and
Matplotlib environment \cite{Oliphant:2007p25672,Hunter:2007p25692}.

\subsection*{Estimating the recombination rate}
To estimate recombination rates, we calculated the frequency of generating the
missing haplotype as a function of the distance between polymorphic sites.
Specifically, the algorithm proceeds as follows: For each pair of biallelic and
gapless sites, we constructed a list of haplotypes, i.e. a list of alleles that
co-occur in the same sequence. The list typically contains 3 or 4 haplotypes,
but can also contain only 2 of the 4 possible pairs due to undersampling or selection
against some of the allele combinations. Only those pairs with 3
haplotypes were included in the estimation. Furthermore, we used only those
sites where both alleles were observed at least three times, since rare alleles
are very sensitive to sampling noise. The analysis was repeated with this
cutoff at two or four alleles, yielding similar results.

For each pair of biallelic sites for which 3 haplotypes were detected in the
sample in timeslice $t_i$, we determined the 4th haplotype that can be formed
from the alleles observed at time $t_i$, i.e.~\base{C\ldots G} in the example
given in Figure \ref{fig:hapgeneration}. We then checked whether this missing haplotype is
detected in timeslice $t_{i+1}$. By averaging over all time points (but the last
one), patients and pairs within a given distance interval, we determined the
frequency of finding the missing haplotype as a function of the distance between
the sites and as a function of the product of distance and time difference. The
error bars indicated in figure \ref{fig:recombination_rate} are calculated as the
product of the estimated value and $1\pm \#  \mathrm{counts}^{-0.5}$. This would
be $\pm$ one standard deviation if all counts were independent, which they are
not since the observations are pairs of sites and each site contributes
to multiple pairs. However, these error bars indicate the relative uncertainties
of the different points, which is all that is needed for an unbiased fit. 
According to eq.~\ref{eq:prob_finding}, we can estimate the recombination rate
from the axis intercept  and the slope at $d\Delta t = 0$, which are extracted
from the data by fitting
\begin{equation}
f(d\Delta t)= c_0+c_1(1-e^{-c_2 d\Delta t})
\end{equation}
to the data. The fit is done by minimizing the squared deviation, weighted by the
relative uncertainty of the data points indicated by the error bars in figure
\ref{fig:recombination_rate}. $\langle Mp_Ap_B\rangle$ was
averaged over all pairs of sites contributing. To estimate a confidence interval 
for the estimate of $\hat{\rho}$, the estimation was repeated 500 times with a set of 11
patients sampled with replacement from the original 11 patients.

The method relies on pairs of biallelic sites in a sample of size $M$, where
3 out of the 4 possible haplotypes are observed. In large samples from a
recombining population, one would naively expect to observe all possible haplotypes. However,
due to the very skewed allele frequency distributions $\sim \frac{1}{p(1-p)}$,
the haplotype formed by the rare alleles is often sufficiently rare that it is
missed even in large samples.  We denote the alleles at site 1 by $a/A$ and at
site 2 by $b/B$, with the $A$ and $B$ being the minor alleles. First, observe
that the mean frequency of haplotype $AB$, $\langle p_{AB}\rangle$, at linkage
equilibrium is $\langle p_{A}\rangle \langle p_{B}\rangle \approx \frac{\log^2
2}{\log^2 M}$. Hence $M \langle p_{AB}\rangle\approx 1$ for $M\leq 20$, in accord
with $M\langle p_{A}\rangle \langle p_{B}\rangle\approx 0.93$ observed in the
data. We further assumed that the frequency of the unobserved haplotype,
$p_{AB}^0$, is significantly smaller than $M^{-1}$. There are several reasons why $p_{AB}^0$
is typically significantly smaller than $\langle p_{A}\rangle \langle
p_{B}\rangle$ and hence smaller than $M^{-1}$: (i) the condition that haplotype
$AB$ is not observed pushes $p_{AB}^0$ down, (ii) minor alleles tend to be in
negative linkage equilibrium if they are involved in selective sweeps, (iii)
allele frequency spectra are even more skewed than $\sim \frac{1}{p(1-p)}$ due to
purifying selection. The degree to which these 3 reasons reduce $p_{AB}^0$ can
be best observed in figure \ref{fig:recombination_rate}B at small $d\Delta t$.
The figure shows the probability to observe haplotype $AB$ at time $t_{i+1}$, given
it was not observed at time $t_i$. At small $d\Delta t$, linkage disequilibrium
is not yet broken down by recombination and the probability to observe
haplotype $AB$ at time $t_{i+1}$ gives us a measure of $Mp_{AB}$ at time
$t_{i}$, which is indeed much smaller than 1. 
Hence we can expand $\exp(-Mp_{AB})$ for small $d\Delta t$ in
Eq.~\ref{eq:prob_finding}.

\subsection*{Estimating the distribution of selection coefficients}
The distribution of the rate of change of allele frequencies was estimated from
pairs of successive time slices with the following characteristics: (i) site
biallelic without indels at time $t_i$ with no constraint on timeslice $t_{i+1}$, and
(ii) monomorphic sites at time $t_i$ that are biallelic at time $t_{i+1}$ (without
indels). In both cases there are two alleles $a$ and $A$ whose frequencies can
be meaningfully compared. The difference in allele frequencies were calculated
as $\Delta p_a = n_a(t_{i+1})/M(t_{i+1})-n_a(t_i)/M(t_i)$, where $n_a(t_i)$ is the
count of allele $a$ at time $t_i$ and $M(t_i)$ is the sample size. Unless a
third allele arose between $t_i$ and $t_{i+1}$ (case (i)), $\Delta p_a=-\Delta
p_A$ and the common estimate of the rate of change, $|\Delta p_a|/\Delta t$, was
added to the cumulative distribution. In rare cases where a third allele did arise, both
$|\Delta p_a|/\Delta t$ and $|\Delta p_A|/\Delta t$ have been added. Sometimes,
a nucleotide changed from one state to another between time $t_i$ or $t_{i+1}$ with no
polymorphisms detected. In such a case, $\Delta p=1$ was added.

To detect the action of selection, we produced cumulative histograms of the rate
of change in allele frequency for nonsynonymous and synonymous (putatively
neutral) polymorphisms by averaging over all patients and pairs of consecutive
time points and $t_i$ and $t_{i+1}$ with both $M(t_i)$ and $M(t_{i+1})$ 
greater than $10$. We then tested for excess of fast changes among the
nonsynonymous polymorphisms using the Kolmogorov-Smirnov test. The test statistic
is the maximal vertical difference between the cumulative distributions $D$
divided by $\sqrt{1/m_{s}+1/m_{n}}$, where $m_{s}$ and $m_{n}$ are the number of
observations of synonymous and non-synonymous polymorphims. The
Kolmogorov-Smirnov test was performed using the statistics module of SciPy
\cite{SciPy}.

To assess the role of hitch-hiking of synonymous polymorphisms with nearby
selected non-synonymous polymorphisms, we produced a list of positions of
non-synonymous polymorphisms whose allele freqency changes faster than $0.002$
per generation between time $t_i$ and $t_{i+1}$. Synonymous polymorphism
closer than 100bp to any of these position where excluded from the
conditioned histogram.

\section*{Acknowledgements} We are grateful to Boris Shraiman  and Erwin Neher for
helpful discussions and would like to thank two anonymous reviewers for 
constructive criticism. We also profited from discussion with
participants of the Kavli Institute for Theoretical Physics on population genetics 
and genomics.  
This work was supported by the National Science Foundation through Grant
PHY05-51164, a LANL LDRD-DR grant (X9R8), and a Harvey L.~Karp Discovery
Award to RAN.

\bibliography{hiv_ms}

\end{document}